\documentstyle[12pt,a41,rotating,amstex,cite,portland]{article}

\newcommand{\Li}{{\rm Li}}

\newcommand{\bq}{\begin{equation}}
\newcommand{\eq}{\end{equation}}
\newcommand\beq{\begin{equation}}
\newcommand\eeq{\end{equation}}
\newcommand\bea{\begin{eqnarray}}
\newcommand\eea{\end{eqnarray}}

\newcommand\Mvec{\,\mbox{\bf M}}

\begin{document}
\setlength{\baselineskip}{0.515cm}
\sloppy
\thispagestyle{empty}
\begin{flushleft}
DESY 08--206 
\\
SFB-CPP/09--002\\
January 2009\\
\end{flushleft}

\mbox{}
\vspace*{\fill}
\begin{center}

{\LARGE\bf Structural Relations of Harmonic Sums and}\\

\vspace*{3mm}
{\LARGE\bf \boldmath Mellin Transforms at  Weight  $w = 
6$~\footnote{Proceedings of the ``Motives, Quantum Field 
Theory, and Pseudodifferential Operators'', held at the Clay Mathematics 
Institute, Boston University, June 2--14, 2008}}\\

\vspace{4cm}
\large
Johannes Bl\"umlein

\vspace{1.5cm}
\normalsize
{\it  Deutsches Elektronen--Synchrotron, DESY,}\\
{\it  Platanenallee 6, D-15735 Zeuthen, Germany}
\\

\end{center}
\normalsize
\vspace{\fill}
\begin{abstract}
\noindent
We derive the structural relations between nested harmonic sums and the 
corresponding Mellin transforms of Nielsen integrals and harmonic polylogarithms
at weight {\sf w = 6}. They emerge in the calculations of massless single--scale 
quantities in QED and QCD, such as anomalous dimensions and Wilson coefficients, to
3-- and 4--loop order. We consider the set of the multiple harmonic sums at weight
six without index $\{-1\}$. This restriction is sufficient for all known physical 
cases. The structural relations supplement the algebraic relations, due to the 
shuffle product between harmonic sums, studied earlier. The original amount of 486 
possible harmonic sums contributing at weight {\sf w = 6} reduces to 99 sums with 
no index $\{-1\}$. Algebraic and structural relations lead to a further reduction 
to 20 basic functions. These functions supplement the set of 15 basic functions up
to weight {\sf w~=~5} derived formerly. We line out an algorithm to obtain the 
analytic representation of the basic sums in the complex plane. 
\end{abstract}

\vspace{1mm}
\noindent

\vspace*{\fill}
\noindent
\numberwithin{equation}{section}
\newpage
\section{Introduction}
%

\vspace{1mm}
\noindent
Inclusive and semi-inclusive scattering cross sections in Quantum Field 
Theories as Quantum Electrodynamics (QED) and Quantum Chromodynamics (QCD)
at higher loop order can be expressed in terms special classes of 
fundamental numbers and functions. Zero scale quantities, like the loop-expansion 
coefficients for renormalized couplings and masses in massless filed theories,
are given by special numbers, which are the multiple $\zeta$-values 
\cite{MZV,HOFPAG}
in the known orders. At higher orders and in the massive case other quantities 
more will contribute 
\cite{ANDRE}. The next class of interest are the single scale quantities to 
which the anomalous dimensions and Wilson coefficients do belong 
\cite{ANDIM,MOMENTS,WILS}, likewise other hard scattering cross sections being 
differential in one variable $z = \hat{L}/L$ given by the ratio of two Lorentz 
invariants with support $z \in [0,1]$. A natural way to study these quantities
consists in representing them in Mellin-space performing the integral 
transform
\begin{equation}
\Mvec\left[f(z)\right](N) = \int_0^1 dz~z^N~f(z)~.
\end{equation}
In the light-cone expansion \cite{LCE} these quantities naturally emerge
as moments for physical reasons with $N \in {\bf N}$. Their mathematical 
representation is obtained 
in terms of nested harmonic sums \cite{HSUM1,HSUM2,HSUM3} 
\begin{equation}
S_{b,\vec{a}}(N) = \sum_{k=1}^N \frac{({\rm sign} b)^k}{k^{|b|}} 
S_{\vec{a}}(k),~~~~S_0(k) = 1~,
\end{equation}
which form a unified language.
This is the main reason to adopt this prescription also for other quantities
of this kind. The harmonic sums lead to the multiple $\zeta$-values 
in the limit $N \rightarrow \infty$ for $b \neq 1$. In the latter case
the harmonic sums diverge.~\footnote{Due to the algebraic relations 
\cite{ALGEBRA} of 
the harmonic sums one may show that this divergence is at most of $O(\ln^m(N))$, 
where $m$ is the number of indices equal to one at the beginning of the index 
set.}
To obtain a representation which is as compact as possible we seek to find
all relations between the harmonic sums. There are two classes of 
relations~:\\
$i)$ the algebraic relations, cf. \cite{ALGEBRA}. They are due to the index 
set of the harmonic sums only and result from their quasi--shuffle algebra 
\cite{HOFM}.\\
$ii)$ the structural relations. These relations depend on the other 
properties of the harmonic sums. One sub--class refers to relations being 
obtained considering harmonic 
sums at $N$ and integer multiples or fractions of $N$, which leads to a 
continuation of $N \in {\bf Q}$. Harmonic sums can be represented
in terms of Mellin-integrals of harmonic polylogarithms $H_{\vec{a}}(z)$
weighted by $1/(1 \pm z)$ \cite{VR}, which belong to the 
Poincar\'{e}--iterated integrals \cite{POIN}.~\footnote{Generalized 
polylogarithms and 
$Z$-sums were considered in \cite{GP}.}
The Mellin integrals are valid for $N \in {\bf R}, N \geq N_0$.
From these representations 
integration-by-parts relations can be derived. Furthermore, there is a large 
number of differentiation relations
\begin{equation}
\frac{d^l}{dN^l} \Mvec\left[f(z)\right](N) 
= \Mvec\left[\ln^l(z) f(z)\right](N)~.
\end{equation}
We analyzed a wide class of physical single scale massless processes and
those containing a single mass scale at two- and three loops 
\cite{ANDIM,MOMENTS,WILS} in 
the past, which led to the same set of {\sf basic harmonic sums} and, related
to it, {\sf basic Mellin transforms}. Like in the case of zero scale 
quantities, this points to a unique representation, which is widely process 
independent and rather related to the contributing Feynman integrals only. The 
representation in terms of harmonic sums is usually more compact than a 
corresponding representation by harmonic polylogarithms, since $i)$ Mellin
convolutions emerge as simple products; $ii)$ harmonic polylogarithms are
multiple integrals, which are usually not reduced to more compact 
analytic representations. The latter requires to solve (part of) these 
integrals analytically.
In the case of harmonic sums the analytic continuation of their argument $N$ 
to complex values has to be performed to apply them in physics problems. As 
lined out in 
Ref.~\cite{DIS07,LL08,STRUCT5} this is possible since harmonic sums can be 
represented in terms of factorial series \cite{FACSER} up to known algebraic 
terms. Harmonic sums turn out to be meromorphic functions with single 
poles at the non-positive integers. One may derive their asymptotic 
representation analytically and they obey recursion relations for complex 
arguments $N$. Due to this their unique representation is given in the complex 
plane.

In the present paper we derive the structural relations of the weight {\sf 
w = 6} harmonic sums extending earlier work on the structural 
relations of harmonic sums up to weight {\sf w = 5} \cite{STRUCT5}.
The paper is organized as follows. In Sections~2--6 we derive the structural 
relations of the harmonic sums of weight {\sf w = 6} of depth 2 to 6 for the 
harmonic sums not containing the index $\{-1\}$. The restriction to this class
of functions is valid in the massless case at least to three-loop order and 
the massive case to two-loop order. In Section~7 we summarize the set of basic 
functions chosen. The principal method to derive the analytic continuation of  
the harmonic sums to complex values of $N$ is outlined in Section~8 in an 
example. Section~9 contains the conclusions. Some useful integrals are 
summarized in the appendix.

\section{Twofold Sums}
%

\vspace{1mm}
\noindent
The following {\sf w = 6} two--fold sums occur~:~~$S_{\pm 5,1}(N), 
S_{\pm 4, \pm 2}(N), 
S_{-3,3}(N)$ and $S_{3,3}(N), S_{-3,-3}(N)$. The latter sums are related 
to single harmonic sums through Euler's relation. 
\begin{equation}
S_{a,b}(N) + S_{b,a}(N) = S_{a}(N) S_{b}(N) + S_{a \wedge b} (N)~,
\end{equation}
with $a \wedge b = {\rm sign}(a) \cdot {\rm sign}(b)(|a| + |b|)$.
For the former six sums we only consider the algebraically irreducible cases. 
In Ref.~\cite{STRUCT5} the basic functions, which determine the 
harmonic sums without index $\{-1\}$ through their Mellin transform, up to 
{\sf w = 5} were found~:
\begin{alignat}{2}
\label{eqBAS1}
{\sf w=1:}~~~& 1/(x-1)
\\
{\sf w=2:}~~~ & \ln(1+x)/(x+1)
\\
{\sf w=3:}~~~ & \Li_2(x)/(x \pm 1)
\\
{\sf w=4:}~~~ & \Li_3(x)/(x + 1), \hspace*{7mm} S_{1,2}(x)/(x \pm 1)
\\
\label{eqBAS2}
{\sf w=5:}~~~ & \Li_4(x)/(x \pm 1), \hspace*{7mm}  S_{1,3}(x)/(x \pm 1),  
\hspace*{7mm}
              S_{2,2}(x)/(x \pm 1), \nonumber\\ & 
 \Li_2^2(x)/(x \pm 1), \hspace*{7mm}
               [\ln(x) S_{1,2}(-x) - \Li^2_2(-x)/2]/(x \pm 1)                     
\end{alignat}
In the following we determine the corresponding basic functions for {\sf w = 6}.

In case of the double sums we show that they all can be related to 
\begin{equation}
\Mvec\left[\frac{\Li_5(x)}{1+x}\right](N)
\end{equation}
up to derivatives of basic functions of lower degree and polynomials of
known harmonic sums.
The representation 
of $S_{\pm 5,1}(N)$ read~:
\begin{eqnarray}
S_{5,1}(N) &=& 
\Mvec\left[\left(\frac{\Li_5(x)}{x-1}\right)_+\right](N) 
- S_1(N) \zeta_5 + S_2(N) \zeta_4 - S_3(N) \zeta_3
               + S_4 \zeta_2  
\\
S_{-5,1}(N) &=& 
(-1)^N \Mvec\left[\frac{\Li_5(x)}{1+x}\right](N) + \frac{15}{16} \zeta_5 
\ln(2) - s_6 
 - S_{-1}(N) \zeta_5 + S_{-2}(N) \zeta_4 
\nonumber\\
& &- S_{-3}(N) \zeta_3
               + S_{-4} \zeta_2~,
\end{eqnarray}
with 
\begin{eqnarray}
\int_0^1 dx g(x) [f(x)]_+ = \int_0^1 dx [g(x) - g(1)] f(x) 
\end{eqnarray}
and
\begin{eqnarray}
s_6 &=& \frac{15}{16} \ln(2) \zeta_5 + \int_0^1 dz \frac{\Li_5(z)}{1+z}
\end{eqnarray}
being one of the basic constants at weight {\sf w} $=6$. For the determination
of the constants in the alternating case we use the tables associated to
Ref.~\cite{HSUM3}. To express one of the sums given below we also give a 
second representation of $S_{-5,1}(N)$,
\begin{eqnarray}
\label{eqSM51}
S_{-5,1}(N) &=& S_{-5}(N) S_1(N) +S_{-6}(N) 
\nonumber\\ & &
+ (-1)^{(N+1)} \Mvec\left[\frac{\Li_5(-x) - \ln(x) \Li_4(-x) + \ln^2(x) 
\Li_3(-x)/2}{x+1}\right](N) \nonumber
\end{eqnarray} \begin{eqnarray}
& &
+ (-1)^{(N+1)} \Mvec\left[\frac{- \ln^3(x) \Li_2(-x)/6 - \ln^4(x) 
\ln(1+x)/24}{x+1}\right](N) \nonumber\\ & &
-\frac{15}{16} \zeta_5 \left[S_{-1}(N) - S_1(N)\right] 
- \frac{23}{70} \zeta_2^3 + \frac{3}{4} \zeta_3^2 + \frac{23}{8}
\zeta_5 \ln(2) - s_6~. 
\end{eqnarray}
The other two-fold sums are
\begin{eqnarray}
S_{-4,-2}(N) &=& - \Mvec\left[\left(\frac{4 \Li_5(-x) - \ln(x) 
\Li_4(-x)}{x-1}\right)_+\right](N) \nonumber\\ & &
+ \frac{1}{2} \zeta_2 \left[S_4(N) -S_{-4}(N)\right] - \frac{3}{2} \zeta_3 
S_3(N) + \frac{21}{8} \zeta_4 S_2(N) - \frac{15}{4} \zeta_5 S_1(N)
\\
S_{-4, 2}(N) &=& (-1)^{N} \Mvec\left[\frac{4 \Li_5(x)-\Li_4(x) \ln(x)}
{1+x}\right](N) \nonumber\\ & &
+ 2 \zeta_3 S_{-3}(N) - 3 \zeta_4 S_{-2}(N) + 4 \zeta_5 S_{-1}(N) 
+ \frac{239}{840} \zeta_2^3 - \frac{3}{4} \zeta_3^2 - \frac{15}{4} \zeta_5 
\ln(2) +4 s_6 
\nonumber\\
\\
S_{ 4,-2}(N) &=& 
\frac{1}{2} \zeta_2 \left[S_{-4}(N)-S_4(N)\right] - \frac{3}{2} \zeta_3
S_{-3}(N) + \frac{21}{8} \zeta_4 S_{-2}(N) - \frac{15}{4} \zeta_5 
S_{-1}(N) 
\nonumber\\ & &
+ (-1)^{N+1} \Mvec\left[\frac{4 \Li_5(-x) - \ln(x) 
\Li_4(-x)}{1+x}\right](N) \nonumber\\ & &
- \frac{313}{840} \zeta_2^3 + \frac{15}{16} \zeta_3^2 + 4 \zeta_5 \ln(2) - 
4 s_6 
\\
S_{ 4, 2}(N) &=& - \Mvec\left[\left(\frac{4 \Li_5(x) - \ln(x) 
\Li_4(x)}{x-1}\right)_+\right](N)
+ 2 \zeta_3 S_3(N) - 3 \zeta_4 S_2(N) +4 \zeta_5 S_1(N)
\nonumber\\
\end{eqnarray}\begin{eqnarray}
S_{-3,-3}(N) &=& 
\Mvec\left[\frac{6 \Li_5(-x) - 3 \ln(x) \Li_4(-x) + \ln^2(x) 
\Li_3(-x)/2}{x-1} \right](N) \nonumber\\ & &
- \frac{3}{4} \zeta_3 \left[S_{-3}(N) - S_3(N)\right] - \frac{21}{8} \zeta_4 S_2(N)
+ \frac{45}{8} \zeta_5 S_1(N) \nonumber\\ 
&=&
\frac{1}{2}  \left[S_{-3}^2(N) + S_6(N)\right] 
\\ 
S_{-3, 3}(N) &=& 3 \zeta_4 S_{-2}(N) - 6 \zeta_5 S_{-1}(N) 
+ (-1)^{N+1} 
6 \Mvec\left[\left(\frac{S_{1,4}(1-x) - \zeta_5}{1+x}\right)_+\right](N) 
\nonumber 
\end{eqnarray}\begin{eqnarray}
& &
+(-1)^{N+1}  
\Mvec\left[
\left(
\frac{
 3\ln(x)  \left[S_{1,3}(1-x) - \zeta_4\right] 
+ \ln^2(x)\left[S_{1,2}(1-x)-\zeta_3\right]/2}
{1+x}\
\right)_+ 
\right](N) 
\nonumber\\ & &
- \frac{271}{280} \zeta_2^3+\frac{81}{32} \zeta_3^2 + \frac{45}{8} \zeta_5 
\ln(2) - 6 s_6 
\\
S_{ 3, 3}(N) &=& 
3 \zeta_4 S_2(N) - 6 \zeta_5 S_1(N) 
- 6\Mvec\left[\left(\frac{S_{1,4}(1-x)-\zeta_5}{x-1}\right)_+\right](N)
\nonumber\\ & &
- \Mvec\left[\left(\frac{\ln^2(x)\left[S_{1,2}(1-x)-\zeta_3\right]/2
+ 3 \ln(x) \left[S_{1,3}(1-x) -\zeta_4\right]}{x-1}\right)_+\right](N)
\nonumber\\ 
&=&
\frac{1}{2}  \left[S_3^2(N) + S_6(N)\right] ~.
\end{eqnarray}
In the above relations Nielsen integrals, \cite{NIELS1}, given by 
\begin{eqnarray}
\label{NIEL1}
S_{p,n}(x) = \frac{(-1)^{p+n+1}}{(p-1)! n!}\int_0^1 \frac{dz}{z} 
\ln^{p-1}(z) \ln^n(1-xz)
\end{eqnarray}
occur. The corresponding functions $S_{1,k}(1-x)$ are given by 
\begin{eqnarray}
S_{1,2}(1-x) &=& - \Li_3(x) + \log(x) \Li_2(x) + \frac{1}{2} \log(1-x)
\log^2(x) + \zeta(3)
\nonumber\\ 
S_{1,3}(1-x) &=& - \Li_4(x) + \log(x) \Li_3(x) - \frac{1}{2} \log^2(x)
\Li_2(x) -\frac{1}{6} \log^3(x) \log(1-x) + \zeta(4)
\nonumber\\ 
S_{1,4}(1-x) &=& - \Li_5(x)+ \ln(x) \Li_4(x) - \frac{1}{2} \ln^2(x) \Li_3(x)  
+ \frac{1}{6} \ln^3(x) \Li_2(x) \nonumber\\  & &+ \frac{1}{24} \ln^4(x)  \ln(1 - x) 
+ \zeta_5~. 
\end{eqnarray}
They are used to express the respective sums in terms of the Mellin 
transforms of basic functions and their derivatives w.r.t. $N$. 

The algebraic relation for $S_{3,3}(N)$ can be used to express
$\Mvec[(\Li_5(x)/(x-1))_+](N)$. The Mellin transform in $S_{-3,-3}(N)$
allows to express $S_{-4,-2}(N)$ and $S_{-4,2}(N)$ through (\ref{eqSM51}).
$S_{4,2}(N)$ and $S_{-3,3}(N)$ do not contain new Mellin transforms. 
Therefore the only non-trivial Mellin transform needed to express the 
double sums at {\sf w = 6} is $\Mvec[\Li_5(x)/(1+x)](N)$. 

In some of the harmonic sums Mellin transforms of the type
\begin{eqnarray}
\frac{\Li_k(-x)}{x \pm 1}~.
\end{eqnarray}
contribute. For odd values of $k = 2 l+1$ the harmonic sums $S_{1,-(k-1)}(N), 
S_{-(k-1),1}(N)$ and $S_{-l,-l}(N)$ allow to substitute the Mellin 
transforms of these functions in terms of Mellin transforms of basic 
functions and derivatives thereof. 

For even values of $k$ this argument applies to $\Mvec[\Li_k(-x)/(1+x)](N)$ 
but not to $\Mvec[\Li_k(-x)/(1+x)](N)$. In the latter case one may use the
relation
\begin{eqnarray}
\frac{1}{2^{k-2}} \frac{\Li_k(x^2)}{1-x^2}
= 
 \frac{\Li_k(x)}{1-x}
+\frac{\Li_k(x)}{1+x}
+\frac{\Li_k(-x)}{1-x}
+\frac{\Li_k(x)}{1+x}~.
\end{eqnarray}
Since in massless quantum field-theoretic calculations both denominators 
occur, one may apply this decomposition based on the first two cyclotomic 
polynomials, cf. \cite{LANG}, and the relation between 
$\Li_k(x^2)$ and $\Li_k(\pm x)$,~\cite{LEWIN}. The corresponding Mellin 
transforms also require half--integer arguments. In more general situations 
other cyclotomic polynomials might emerge.   
The relation
\begin{eqnarray}
\label{eqREL1}
\frac{1}{2^{k-1}} 
\Mvec\left[\left(\frac{\Li_k(x^2)}{x^2-1}\right)_+\right]
\left(\frac{N-1}{2}\right)
&=& 
 \Mvec\left[\left(\frac{\Li_k(x)}{x-1}\right)_+\right](N)
+\Mvec\left[\left(\frac{\Li_k(x)}{x+1}\right)_+\right](N)
\nonumber\\ & &
+\Mvec\left[\left(\frac{\Li_k(-x)}{x-1}\right)_+\right](N)
+\Mvec\left[\left(\frac{\Li_k(-x)}{x+1}\right)_+\right](N)
\nonumber\\ & &
- \int_0^1 dx \frac{\Li_k(x^2)}{1+x}
\end{eqnarray}
determines $\Mvec[\Li_k(-x)/(1+x)](N)$. For $k = 2,4$  the last integral in
(\ref{eqREL1}) is given by
\begin{eqnarray}
\int_0^1 dx \frac{\Li_2(x^2)}{1+x} &=& \zeta_2 \ln(2) - \frac{3}{4} 
\zeta_3\\
\int_0^1 dx \frac{\Li_4(x^2)}{1+x} &=& \frac{2}{5} \ln(2) \zeta_2^2 + 3 
\zeta_2 \zeta_3 - \frac{25}{4} \zeta_5~.
\end{eqnarray}
The corresponding relations for $\Mvec[\Li_k(-x)/(1+x)](N)$ are~:
\begin{eqnarray}
\Mvec\left[\frac{\Li_2(-x)}{x+1}\right](N) &=& - \frac{1}{2} 
\Mvec\left[\left(\frac{\Li_2(x)}{x-1}\right)_+\right]\left(\frac{N-1}{2}\right)
+\Mvec\left[\left(\frac{\Li_2(x)}{x-1}\right)_+\right]\left( N \right)
\nonumber\\ & &
+\Mvec\left[\left(\frac{\Li_2(-x)}{x-1}\right)_+\right]\left( N \right)
-\Mvec\left[\frac{\Li_2(x)}{x+1}\right]\left( N \right)
\nonumber\\ & &
+\frac{3}{8} \zeta_3 - \frac{1}{2} \zeta_2 \ln(2)
\\
\Mvec\left[\frac{\Li_4(-x)}{x+1}\right](N) &=&  
- \frac{1}{8} 
\Mvec\left[\left(\frac{\Li_4(x)}{x-1}\right)_+\right]\left(\frac{N-1}{2}\right)
+\Mvec\left[\left(\frac{\Li_4(x)}{x-1}\right)_+\right]\left( N \right)
\nonumber\\ & &
+\Mvec\left[\left(\frac{\Li_4(-x)}{x-1}\right)_+\right]\left( N \right)
-\Mvec\left[\frac{\Li_4(x)}{x+1}\right]\left( N \right)
\nonumber\\ & &
-\frac{1}{20} \zeta_2^2 \ln(2) - \frac{3}{8} \zeta_2 \zeta_3
+\frac{25}{32} \zeta_5~.
\end{eqnarray}
In the case of {\sf w = 6} these relations do not lead to a further reduction
of basic functions but are required at lower weights, cf. \cite{STRUCT5}.

\section{Threefold Sums}             
%

\vspace{1mm}
\noindent
The triple sums are~:
\begin{eqnarray}
S_{4,1,1}(N) &=&
- \Mvec\left[\left(\frac{S_{3,2}(x)}{x-1}\right)_+\right](N) 
+ S_1(N) (2 \zeta_5 - \zeta_2 \zeta_3) - 
\frac{\zeta_4}{4} S_2(N) + \zeta_3 S_3(N) 
\\
S_{-4,1,1}(N) &=& 
(-1)^{N+1} \Mvec\left[\frac{S_{3,2}(x)}{1+x}\right](N) 
+(2 \zeta_5 - \zeta_2 \zeta_3) S_{-1}(N)  
- \frac{\zeta_4}{4} S_{-2}(N)  
\nonumber
\\ 
& & 
+ \zeta_3 S_{-3}(N) 
+ \frac{71}{840} \zeta_2^3 +\frac{1}{8} \zeta_3^2
- \frac{29}{32} \zeta_5 \ln(2) - \zeta_2 \zeta_3 \ln(2) +\frac{3}{2} s_6
\\
S_{-3,-2,1}(N) &=&  
\Mvec\left[\frac{H_{0,0,-1,0,1}(x)}{x-1}\right](N) + \zeta_2 S_{-3,-1}(N)
+\left[S_{-3}(N) - S_3(N)\right] \left[\zeta_2 \ln(2) - \frac{5}{8} 
\zeta_3\right] 
\nonumber\\ &&
+ \frac{3}{40} \zeta_2^2 S_2(N) - \left(\frac{9}{4} \zeta_2 
\zeta_3 - \frac{67}{16} \zeta_5\right) S_1(N)
\\
S_{-2,-3,1}(N) &=&
S_{-2}(N) S_{-3,1}(N) +S_{5,1}(N) + S_{-3,-3}(N) - S_{-3,1,-2} - S_{-3,-2,1}
\\
S_{1,-2,-3}(N) &=&  S_{-3}(N) S_{1,-2}(N) + S_{1,5}(N)
- S_1(N) S_{-3,-2}(N)- S_{-3,-3}(N) 
\nonumber\\ &&
+ S_{-3,-2,1}(N)  
\\
S_{1,-3,-2}(N) &=& 
S_1(N) S_{-3,-2}(N) + S_{-4,-2}(N) + S_{-3,-3}(N) - S_{-3,1,-2}(N) - 
S_{-3,-2,1}(N)
\\
S_{-2,1,-3}(N) &=& 
S_{-3,-3}(N) - S_{-3}(N) S_{1,-2}(N) - S_{1,5}(N) + S_1(N) S_{-2,-3}(N)
\nonumber\\ &&
- S_{-2}(N) S_{-3,1}(N) - S_{5,1}(N) + S_{-2,-4}(N) +S_1(N) S_{-3,-2}(N)
\nonumber\\ &&
+S_{-3,1,-2}(N)
\\
S_{-3,1,-2}(N) &=&
\Mvec\left[\left(\frac{A_1(-x)/2+S_{3,2}(-x)-S_{2,2}(-x)\ln(x)}{x-1}\right)_+\right](N)
\nonumber\\ & &
-\frac{1}{2} \zeta_2 \left[S_{-3,1}(N)-S_{-3,-1}(N)\right]
-\left[\frac{1}{8} \zeta_3 - \frac{1}{2} \zeta_2 \ln(2)\right]\left[S_{-3}(N)-S_3(N)\right]
\nonumber\\ & & + \frac{1}{8} \zeta_2^2 S_2(N) + \left[\frac{23}{16} \zeta_5 - \frac{7}{8} \zeta_2 
\zeta_3\right] S_1(N) 
\\ 
S_{-3,2,1}(N) &=& (-1)^N \Mvec\left[\frac{2 S_{3,2}(x) - 
A_1(x)/2}{x+1}\right](N) 
+ \zeta_2 S_{-3,1}(N) - \frac{3}{4} \zeta_4 S_{-2}(N) 
\nonumber\\ &&
- \left(\frac{11}{2} \zeta_5 - 3 \zeta_2 \zeta_3 \right) S_{-1}(N)
+ \frac{23}{168} \zeta_2^3 + \frac{59}{64} \zeta_3^2
+ \frac{41}{32} \zeta_5 \ln(2) + \frac{1}{2} \zeta_2^2 \ln^2(2) 
\nonumber\\ &&
+ \frac{5}{4} \zeta_2 \zeta_3 \ln(2) - \frac{1}{12} \zeta_2 \ln^4(2)
- 2 \zeta_2 \Li_4\left(\frac{1}{2}\right)
- \frac{7}{2} s_6 
\\
S_{2,-3,1}(N) &=&  (-1)^N \Mvec\left[\frac{H_{0,-1,0,0,1}(x)}{1+x}\right](N)
- \left(\frac{83}{16} \zeta_5 - 
\frac{21}{8} \zeta_2 \zeta_3\right) S_{-1}(N)
\nonumber\\
&& 
+ \zeta_2 S_{2,-2}(N) - \zeta_3 S_{2,-1}(N) 
\nonumber\\
&& 
+\left[-\frac{3}{5} \zeta_2^2  +2 \Li_4\left(\frac{1}{2}\right) + 
\frac{3}{4} \zeta_3 \ln(2) - \frac{1}{2} \zeta_2 \ln^2(2) +\frac{1}{12} 
\ln^4(2) \right] \left[S_2(N) - S_{-2}(N) \right] 
\nonumber\\
&& - \frac{7}{8} \zeta_2 \zeta_3 \ln(2) - \frac{1}{6} \zeta_2 \ln^4(2)
   - 4 \zeta_2 \Li_4\left(\frac{1}{2}\right) + \zeta_2^2 \ln^2(2)
   + \frac{11}{12} \zeta_2^3 + \frac{87}{64} \zeta_3^2 
\nonumber\\
&&
   - \frac{11}{32} \zeta_5 \ln(2) - \frac{5}{2} s_6
\\
S_{1,2,-3}(N) &=& 
S_{-3}(N) S_{1,2}(N) + S_{1,-5}(N) - S_1(N) S_{-3,2}(N) - S_{-3,3}(N) 
+ S_{-3,2,1}(N) 
\\
S_{1,-3,2}(N) &=& - S_2(N) S_{-3,1}(N) - S_{-5,1}(N) + S_{2,-3,1}(N) 
+ S_1(N) S_{-3,2}(N)
\nonumber\\ &&
+ S_{-4,2}(N) 
\end{eqnarray} \begin{eqnarray}
S_{2,1,-3}(N) &=&  S_{3,-3}(N) - S_{-3}(N) S_{1,2}(N) -S_{1,-5}(N)
+ S_1(N) S_{2,-3}(N) - S_{2,-3,1}(N) 
\nonumber\\ &&
+ S_{2,-4}(N) + S_1(N) S_{-3,2}(N)
+ S_{-3,3}(N) - S_{-3,2,1}(N)
\\
S_{-3,1,2}(N) &=& 
S_2(N) S_{-3,1}(N) + S_{-5,1}(N) + S_{-3,3}(N) - S_{2,-3,1}(N) 
- S_{-3,2,1}(N) 
\\
S_{-2,3,1}(N)  &=&
(-1)^N \Mvec\left[\frac{(3/2) A_1(x) - \Li_2(x) 
\Li_3(x)}{x+1}\right](N) 
+ \zeta_2 S_{-2,2}(N) - \zeta_3 S_{-2,1}(N)  \nonumber\\ & &
+ \left( \frac{9}{2} \zeta_5
- 2 \zeta_2 \zeta_3\right) S_{-1}(N)
- \frac{13}{12} \zeta_2^3 - \frac{43}{32} \zeta_3^2 + \frac{51}{32} \zeta_5 
\ln(2) - \zeta_2^2 \ln^2(2) 
\nonumber\
\\ 
&&
+ \frac{3}{2} \zeta_2 \zeta_3 \ln(2) 
+ \frac{1}{6}
\zeta_2 \ln^4(2) + 4 \zeta_2 \Li_4\left(\frac{1}{2}\right) + \frac{3}{2} s_6
\\
S_{3,-2,1}(N) &=&
S_{3,-3}(N) - S_{3,1,-2}(N) - S_{-2,3,1}(N) + S_{-2}(N) S_{3,1}(N)
+ S_{-5,1}(N)
\\ 
S_{1,-2,3}(N)  &=&  S_{-3,3}(N) - S_3(N) S_{-2,1}(N) - S_{-2,4}(N)
+ S_{-2}(N) S_{1,3}(N) + S_{1,-5}(N) 
\nonumber\\ &&
+ 2 S_{-2}(N) S_{3,1}(N)
+ S_{3,-3}(N) - S_1(N) S_{3,-2}(N) - S_{4,-2}(N) + S_{-5,1}(N)
\nonumber\\ &&
- S_{3,1,-2}(N) - S_{-2,3,1}(N)
\\
S_{1,3,-2}(N)  &=& -S_{-2}(N) S_{3,1}(N) - S_{-5,1}(N) + S_{-2,3,1}(N)
                   + S_1(N) S_{3,-2}(N) 
\nonumber\\ & &
+ S_{4,-2}(N)  
\\
S_{-2,1,3}(N)  &=& 
S_3(N) S_{-2,1}(N) + S_{-2,4}(N) - S_{-2}(N) S_{3,1}(N) - S_{3,-3}(N)
+ S_{3,1,-2}(N)
\\
S_{3,1,-2}(N)  &=& 
(-1)^N \Mvec\left[\frac{A_1(-x)/2+S_{3,2}(-x)-\ln(x) S_{2,2}(-x)}
                                {1+x}\right](N)
\nonumber\\ & &
- \frac{1}{2} \zeta_2 \left[S_{3,1}(N) - S_{3,-1}(N) \right] 
- \left[\frac{1}{8} \zeta_3 -\frac{1}{2} \zeta_2 \ln(2) \right] 
  \left[S_3(N) - S_{-3}(N) \right]
\nonumber\\ & &
+ \frac{1}{8} \zeta_2^2 S_{-2}(N) 
+ \left[\frac{23}{16} \zeta_5 - \frac{7}{8} \zeta_2 \zeta_3\right] S_{-1}(N)
\nonumber\\ &&
+ \frac{113}{560} \zeta_2^3 - \frac{17}{64} \zeta_3^2 
- \frac{1}{2} \zeta_5 \ln(2) - \frac{7}{8} \zeta_2 \zeta_3 \ln(2) + s_6
\\
S_{3,2,1}(N)  &=& \Mvec\left[\left(\frac{2 S_{3,2}(x) - 
A_1(x)/2}{x-1}\right)_+\right](N)
\nonumber\\
& & + \zeta_2 S_{3,1}(N) - \frac{3}{4} \zeta_4 S_2(N) 
- \left(\frac{11}{2} \zeta_5 - 3 \zeta_2 \zeta_3 \right) S_1(N)
\\
S_{2,3,1}(N)  &=& \Mvec\left[\left(\frac{(3/2) A_1(x) - \Li_2(x) 
\Li_3(x)}{x-1}\right)_+\right](N) \nonumber\\ & &
+ \zeta_2 S_{2,2}(N) - \zeta_3 S_{2,1}(N) + \left( \frac{9}{2} \zeta_5
- 2 \zeta_2 \zeta_3\right) S_1(N)
\\
S_{1,2,3}(N)  &=& S_3(N) S_{1,2}(N)+S_{1,5}(N) - S_1(N) S_{3,2}(N)
- S_{3,3}(N) + S_{3,2,1}(N) 
\\
S_{2,1,3}(N)  &=& 2 S_{3,3}(N) - S_3(N) S_{1,2}(N) - S_{1,5}(N)
+ S_1(N) S_{2,3}(N) - S_{2,3,1}(N) 
\nonumber\\ & &
+ S_{2,4}(N) 
+ S_1(N) S_{3,2}(N)
- S_{3,2,1}(N)
\\
S_{1,3,2}(N)  &=& - S_2(N) S_{3,1}(N) - S_{5,1}(N) + S_{2,3,1}(N) 
+ S_1(N) S_{3,2}(N) + S_{4,2}(N)
\\
S_{3,1,2}(N)  &=& S_2(N) S_{3,1}(N) + S_{5,1}(N) +S_{3,3}(N) - 
S_{2,3,1}(N)
- S_{3,2,1}(N)
\\
\label{eqttt}
S_{2,2,2}(N)    &=&  -\Mvec\left[\left(\frac{2A_1(x) + \Li_2^2(x) \ln(x)/2
                       - 2 S_{2,2}(x) \ln(x)}{x-1}\right)_+\right](N)
\nonumber\\ & &
                     -\Mvec\left[\left(\frac{4 S_{3,2}(x) - 2 \Li_2(x) \Li_3(x)}{x-1}
\right)_+\right](N)
\nonumber\\ & & 
+ 2\zeta_3 S_{2,1}(N) + 2 \left(\zeta_5 - \zeta_2 \zeta_3\right) S_1(N) 
\nonumber
\end{eqnarray}\begin{eqnarray}
                &=&  \frac{1}{6} S_{2}^3(N) + \frac{1}{2} S_{2}(N) S_4(N)
                    +\frac{1}{3} S_{6}(N)
\\
S_{-2,2,2}(N)    &=&  (-1)^{(N+1)}\Mvec\left[\frac{2A_1(x) + \Li_2^2(x) \ln(x)/2
                       - 2 S_{2,2}(x) \ln(x)}{x+1}\right](N)
\nonumber\\ & &
                     +(-1)^{(N+1)}\Mvec\left[\frac{4 S_{3,2}(x) - 2 \Li_2(x) \Li_3(x)}{x+1}
\right](N)\nonumber\\ & & 
+ 2\zeta_3 S_{-2,1}(N) + 2 \left(\zeta_5 - \zeta_2 \zeta_3\right) S_{-1}(N) 
\nonumber
\\ 
&&
-\frac{1}{4} \zeta_2 \zeta_3 \ln(2) + \frac{1}{12} \zeta_2 \ln^4(2)
+ 2 \zeta_2 \Li\left(\frac{1}{2}\right) - \frac{1}{2} \zeta_2^2 \ln^2(2)
+ \frac{37}{80} \zeta_2^3 - 2 \zeta_3^2 
\nonumber\\ &&
- \frac{23}{4} \zeta_5 \ln(2)
+ 4 s_6
\\
S_{2,-2,2}(N)   &=& - 2 S_{-2,2,2}(N) + S_2(N) S_{-2,2}(N) + S_{-4,2}(N) 
+ S_{-2,4}(N) 
\\
S_{2,2,-2}(N)   &=& S_{-2,2,2}(N) - \frac{1}{2} \left[ S_2(N) S_{-2,2}(N)
+ S_{-4,2}(N) + S_{-2,4}(N) 
\right. 
\nonumber\\ & & \left.
- S_2(N) S_{2,-2}(N)
- S_{4,-2}(N) - S_{2,-4}(N) \right]
\\
S_{-2,2,-2}(N)   &=&
\Mvec\left[\left(\frac{-4 S_{3,2}(-x) - \ln(x) \Li_2^2(-x)/2 + 2 \ln(x) S_{2,2}(-x)}{x-1} \right)_+\right](N)
\nonumber\\ & & 
+ \Mvec\left[\left(\frac{2 \Li_3(-x) \Li_2(-x) - 2 A_1(-x)}{x-1}\right)_+\right](N)
\nonumber\\ & &
- \frac{3}{2} \zeta_3 S_{-2,-1}(N) + \frac{1}{2} \zeta_2\left[S_{-2,-2}(N) - S_{-2,2}(N)\right]
\nonumber\\ & &
+\left[-\frac{11}{8} \zeta_2^2+ 4 \Li_4\left(\frac{1}{2}\right) + 2 \zeta_3 \ln(2)
- \zeta_2 \ln^2(2) + \frac{1}{6} \ln^4(2) \right] \left[S_{-2}(N) -S_2(N)\right]
\nonumber \\
&& + \left(\frac{11}{4} \zeta_2 \zeta_3 - \frac{23}{4} \zeta_5\right) S_1(N)
\\
S_{2,-2,-2}(N)   &=&  \frac{1}{2} \left[-S_{-2,2,-2}(N) + S_{-2}(N) S_{2,-2}(N) 
+ S_{-4,-2}(N) 
                      + S_{2,4}(N) \right] 
\\
S_{-2,-2,2}(N)  &=&  \frac{1}{2}\left[ - S_{-2,2,-2}(N) + S_{-2}(N) S_{-2,2}(N) + S_{4,2}(N)
                     + S_{-2,-4}(N)\right]
\\
S_{-2,-2,-2}(N) &=& \frac{1}{6} S_{-2}^3(N) + 
\frac{1}{2} S_{-2}(N) S_4(N)+\frac{1}{3} S_{-6}(N)~. 
\end{eqnarray}
There emerge numerator functions, which do not belong to the class of Nielsen
integrals~\footnote{Note a
misprint in Eq.~(14),~\cite{LL08}. $\Li_4(y)$ should read $\Li_4(1-y)$.},
\begin{eqnarray}
A_1(x) &=& \int_0^x \frac{dy}{y} \Li_2^2(y) \nonumber\\
A_2(x) &=& \int_0^x \frac{dy}{y} \ln(1-y) S_{1,2}(y) \nonumber\\
A_3(x) &=& \int_0^x \frac{dy}{y} [\Li_4(1-y) -\zeta_4]~.
\end{eqnarray}
As seen in Eqs.~(\ref{eqttt}), $(A_1(x)/(x-1))_+$
is not a basic function since its Mellin transform reduces to 
single harmonic sums and known Mellin transforms algebraically.
Furthermore, some numerator functions are given by harmonic polylogarithms
$H_{a_1, \ldots, a_k}(x),~~a_i~\in~\{-1,0,+1\},$
which cannot be reduced significantly further. Harmonic polylogarithms are
Poincar\'{e}--iterated integrals \cite{POIN} over 
the alphabet $[f_0,f_1,f_{-1}] = [1/x,1/(x-1),1/(x+1)]$,~\cite{VR}, with
\begin{eqnarray}
H_{0}(x) &=& \hspace{4mm} \ln(x) \\
H_{1}(x) &=& -\ln(1-x) \\
H_{-1}(x) &=& \hspace{4mm} \ln(1+x) 
\end{eqnarray}
and
\begin{eqnarray}
H_{a,\vec{b}}(x) &=& \int_0^x dy~f_a(y) H_{\vec{b}}(y)~.
\end{eqnarray}

\section{Fourfold Sums}
%

\vspace{1mm}
\noindent
The  quadruple--index sums are~:
\begin{eqnarray}
S_{-3,1,1,1}(N)   &=& 
(-1)^{N} \Mvec\left[\frac{S_{2,3}(x)}{x+1}\right](N)  
+ \zeta_4 S_{-2}(N) -(2 \zeta_5 -\zeta_2 \zeta_3) S_{-1}(N) \nonumber\\
&&
+ \frac{1}{8} \zeta_2 \zeta_3 \ln(2) - \frac{1}{6} \zeta_2 \ln^4(2)
- \zeta_2 \Li_4\left(\frac{1}{2}\right) + \frac{1}{4} \zeta_2^2 \ln^2(2)
- \frac{257}{840} \zeta_2^3 + \frac{7}{24} \zeta_3 \ln^3(2)
\nonumber\\ &&
+ \frac{41}{64} \zeta_3^2 - \frac{33}{32} \zeta_5 \ln(2) + 2 \ln(2) 
\Li_5\left(\frac{1}{2}\right) + \ln^2(2) \Li_4\left(\frac{1}{2}\right)
\nonumber\\ &&
+\frac{1}{36} \ln^6(2)
+ 2 \Li_6\left(\frac{1}{2}\right) - \frac{s_6}{2}
\\
S_{3,1,1,1}(N)   &=& 
-\Mvec\left[
\left(\frac{S_{2,3}(x)}{x-1}\right)_+\right](N)
+ \zeta_4 S_2(N) -(2 \zeta_5 -\zeta_2 \zeta_3) S_1(N) 
\\
S_{-2, 2,1,1}(N) &=& (-1)^{N+1} \Mvec\left[\frac{3 S_{2,3}(x) + 
A_2(x)}{x+1}\right](N) + \zeta_3 S_{-2,1}(N)
+ \left(\frac{11}{2}\zeta_5-3 \zeta_2 \zeta_3  \right) S_{-1}(N)
\nonumber\\ &&
-\frac{5}{4} \zeta_2 \zeta_3 \ln(2) + \frac{1}{3} \zeta_2 
\ln^4(2) + 2 \zeta_2 \Li_4\left(\frac{1}{2}\right) - 
\frac{1}{2} \zeta_2^2 \ln^2(2) + \frac{411}{560} \zeta_2^3 - \frac{7}{12} 
\zeta_3 \ln^3(2) 
\nonumber\\ &&
- \frac{9}{8} \zeta_3^2 + \frac{73}{64} \zeta_5 \ln(2) 
- 4 \ln(2) \Li_5\left(\frac{1}{2}\right) - 2 \ln^2(2) 
\Li_4\left(\frac{1}{2}\right) - \frac{1}{18} \ln^6(2) 
\nonumber\\ &&
- 4 
\Li_6\left(\frac{1}{2}\right) + \frac{9}{4} s_6
\\
S_{2,-2,1,1}(N) &=&  (-1)^{N+1} \Mvec\left[\frac{H_{0,-1,0,1,1}(x)}{1+x}
\right](N) + \zeta_3 S_{2,-1}(N) + \left[\frac{11}{16} \zeta_2 \zeta_3
- \frac{41}{32} \zeta_5 \right] S_{-1}(N)
\nonumber\\ &&
+ \left[ -\Li_4\left(\frac{1}{2}\right) + \frac{1}{8} \zeta_2^2
+ \frac{1}{8} \zeta_3 \ln(2) + \frac{1}{4} \zeta_2 \ln^2(2) - \frac{1}{24} 
\ln^4(2) \right] \left[S_2(N) - S_{-2}(N)\right]
\nonumber\\ &&
-\frac{17}{16} \zeta_2 \zeta_3 \ln(2) - \frac{1}{3} \zeta_2 \ln^4(2) - 2 
\zeta_2 \Li_4\left(\frac{1}{2}\right) + \frac{1}{2} \zeta_2^2 \ln^2(2) 
- \frac{87}{280} \zeta_2^3 + \frac{7}{12} \zeta_3 \ln^3(2) 
\nonumber\\ &&
+ \frac{105}{128} 
\zeta_3^2 - \frac{103}{32} \zeta_5 \ln(2) + 4 \ln(2) 
\Li_5\left(\frac{1}{2}\right) + 2 \ln^2(2) \Li_4\left(\frac{1}{2}\right) + 
\frac{1}{18} \ln^6(2) 
\nonumber\\ &&
+ 4 \Li_6\left(\frac{1}{2}\right) + s_6
\\
S_{-2, 1,1,2}(N) &=& (-1)^N \Mvec\left[\frac{A_3(x)}{1+x}\right](N)
- \left(\zeta_2 \zeta_3 - 3 \zeta_5\right) S_{-1}(N) - \zeta_3 S_{-2,1}(N)
+ \zeta_2 S_{-2,1,1}(N) 
\nonumber\\ &&
+ \frac{5}{16} \zeta_2 \zeta_3 \ln(2) + \frac{1}{16} \zeta_2 
\ln^4(2) + \frac{3}{2} \zeta_2 \Li_4\left(\frac{1}{2}\right) - \frac{3}{8}
\zeta_2^2 \ln^2(2) + \frac{11}{120} \zeta_2^3 - \frac{27}{16} \zeta_3^2
\nonumber
\end{eqnarray}\begin{eqnarray}
 &&
- \frac{59}{32} \zeta_5 \ln(2) + \frac{5}{2} s_6
\\
S_{-2,-2,1,1}(N) &=& - \Mvec\left[\left(\frac{H_{0,-1,0,1,1}(x)}{x-1}\right)_+
\right](N) + \zeta_3 S_{-2,-1}(N) 
+\left(\frac{11}{16} \zeta_2 \zeta_3 - \frac{41}{32} \zeta_5\right) S_1(N)
\nonumber\\ &&
+ 
\left(- \Li_4\left(\frac{1}{2}\right) + \frac{1}{8} \zeta_2^2 + 
\frac{1}{8} \zeta_3 \ln(2) + \frac{1}{4} \zeta_2 \ln^2(2) - \frac{1}{24} 
\ln^4(2)\right)
\left[S_{-2}(N) - S_2(N)\right]
\nonumber\\	
\\
S_{ 2, 2,1,1}(N) &=& - \Mvec\left[\left(\frac{3 S_{2,3}(x)+A_2(x)}{x-1}
\right)_+\right](N) + \zeta_3 S_{2,1}(N) 
+\left(\frac{11}{2} \zeta_5 -3 \zeta_2 \zeta_3 \right) S_1(N)~.
\nonumber\\
\end{eqnarray}
Here, the harmonic polylogarithm $H_{0,-1,0,1,1}(x)$ is given by
\begin{eqnarray}
H_{0,-1,0,1,1}(x) &=& \int_0^x \frac{dy}{y} \int_0^y dz 
\frac{S_{1,2}(z)}{1+z}~.
\end{eqnarray}
We tested the above sum-relations containing harmonic polylogarithms 
in the Mellin transforms numerically using the code of Ref.~\cite{GR}.
\section{Fivefold Sums}
%

\vspace{1mm}
\noindent
Two  5--fold sums contribute~:
\begin{eqnarray}
S_{2,1,1,1,1}(N) &=&  
- \Mvec\left[\left(\frac{S_{1,4}(x)}{x-1}\right)_+\right](N) +
\zeta_5 S_1(N) 
\\
S_{-2,1,1,1,1}(N) &=& 
(-1)^{N+1} \Mvec\left[\frac{S_{1,4}(x)}{1+x}\right](N) 
+ \zeta_5 S_{-1}(N) \nonumber\\ &&
+ \frac{7}{16}\zeta_2 \zeta_3 \ln(2) 
+ \frac{1}{12} \zeta_2 \ln^4(2) 
+ \frac{1}{2}\zeta_2 \Li_4 \left(\frac{1}{2}\right) 
- \frac{1}{8} \zeta_2^2 \ln^2(2) 
- \frac{7}{48} \zeta_3 \ln^3(2) 
\nonumber\\ & &
- \frac{49}{128} \zeta_3^2 
- \ln(2) \Li_5\left(\frac{1}{2}\right) 
- \frac{1}{2} \ln^2(2) \Li_4\left(\frac{1}{2}\right) 
- \frac{1}{72} \ln^6(2) 
- \Li_6\left(\frac{1}{2}\right) 
+ \ln(2) \zeta_5~.
\nonumber\\
\end{eqnarray}
All other sums can be traced back to these sums using algebraic 
relations \cite{ALGEBRA}. The other Mellin transforms emerging in their 
representation
were all calculated  in Refs.~\cite{HSUM2,STRUCT5} before. 
\section{Sixfold Sums}
%

\vspace{1mm}
\noindent
Only one sixfold sum contributes at {\sf w = 6}, $S_{1,1,1,1,1,1}(N)$.
This sum is completely reducible into a polynomial of single harmonic 
sums, cf.~\cite{HSUM2},
\begin{eqnarray}
S_{\underbrace{\mbox{\scriptsize 1, \ldots ,1}}_{\mbox{\scriptsize
6}}}
&=&
 \frac{1}{720}\,      S_{1}^{6}  
+\frac{1}{48}\,S_{2}\,S_{1}^{4} 
+\frac{1}{18}\,S_{3}\,S_{1}^{3} 
+\frac{1}{8}\, S_{4}  S_{1}^2     
+\frac{1}{5}\, S_{5}\,S_{1}
+\frac{1}{16}\,S_{1}^{2} S_{2}^{2}
\nonumber\\ & &
+\frac{1}{6}\, S_{1}\,S_{2}\,S_{3} 
+\frac{1}{48}\,S_{2}^{3}
+\frac{1}{8}\, S_{2}\,S_{4}
+\frac{1}{18}\,S_{3}^{2}
+\frac{1}{6}\, S_{6} 
\end{eqnarray}
\section{The Basic Functions}
%

\vspace{1mm}
\noindent
In the following we summarize the {\sf basic functions} the Mellin transforms
of which represents the harmonic sums up to weight {\sf w = 6} without those
carrying an index $\{-1\}$. The corresponding sums of lower weight were
determined in Refs.~\cite{HSUM1,JBVR,STRUCT5}. The new 20 functions 
are given by 
\begin{alignat}{2}
{\sf w=6:}~~~ & \Li_5(x)/(x + 1), \hspace*{7mm}
 S_{1,4}(x)/(x \pm 1),  \hspace*{7mm}
              S_{2,3}(x)/(x \pm 1),   \nonumber\\
& S_{3,2}(x)/(x \pm 1), \hspace*{7mm}
               \Li_2(x) \Li_3(x)/(x \pm 1), \hspace*{7mm}
              \nonumber\\
&A_1(x)/(x+1), \hspace*{9mm}
A_2(x)/(x \pm 1), \hspace*{8mm}
A_3(x)/(x+1) \\
&H_{0,-1,0,1,1}(x)/(x \pm 1), \hspace*{7mm} 
 H_{0,0,-1,0,1}(x)/(x \pm 1)
\nonumber\\
&[A_1(-x)+2 S_{3,2}(-x)-2 S_{2,2}(-x) \ln(x) ]/(x \pm 1) \nonumber\\
&[A_1(-x) + 2 S_{3,2}(-x) - S_{2,2}(-x) \ln(x) + \Li_2^2(-x) \ln(x)/4
- \Li_3(-x) \Li_2(-x)]/(x-1)
\nonumber
\end{alignat} 
and extend the set Eqs.~(\ref{eqBAS1}--\ref{eqBAS2}). The algebraic relations
allow to express the initial set of 99 functions by 30 functions and
the structural relations reduce the basis further to 20 functions.

\section{Complex Analysis of Harmonic Sums}
%

\vspace{1mm}
\noindent
The anomalous dimensions and Wilson coefficients expressed in Mellin space
allow simple representations of the scale evolution of single-scale 
observables, which are given by ordinary differential equations. The 
experimental measurement of the observables requires the representation in 
$z$--space. 
Therefore, one has to perform the analytic continuation
of harmonic sums
to complex values of $N$. Precise numerical representations for 
the analytic continuation of the basic functions up to weight {\sf w = 5}
were derived in \cite{ANCONT1} based on the {\tt MINIMAX}-method~\cite{MINIMAX}. 
One may even obtain corresponding representations for quite general functions
$\Phi(z),~~z~\in~[0,1]$, as worked out for the heavy flavor Wilson 
coefficients 
to 2-loop order in \cite{AB}.~\footnote{For another 
proposal for the analytic continuation of harmonic sums to $N~\in~{\bf R}$, 
for which some simple examples were presented, cf. \cite{KOT}.} For other 
effective parameterizations see \cite{Vogt1}. 

Here we aim on an exact representations. The inverse Mellin transforms are 
obtained by a contour integral around the singularities of the respective 
functions in the complex plane.

We traced back all the harmonic sums to Mellin transforms of basic functions
$f_i(z)$,
\begin{eqnarray}
\label{MEL1}
F_i^-(N) = \int_0^1 dz f_i(z) \frac{z^N 
-1}{z-1}, \hspace*{2cm} F_i^+(N) = \int_0^1 dz f_i(z) \frac{(-z)^N 
-1}{z+1}~.
\end{eqnarray}
Eqs.~(\ref{MEL1}) imply the recursion relations
\begin{eqnarray}
F_i^-(N+1) &=& - F_i^-(N) + \int_0^1 dz z^N f_i(z)~,\\ 
F_i^+(N+1) &=& \hspace{3mm}  F_i^+(N) + (-1)^{N+1} \int_0^1 dz z^N f_i(z)~. 
\end{eqnarray}
The remaining integrals are simpler Mellin transforms, which correspond to 
harmonic sums of lower weight.

If the functions $f_i(z)/(z-1),~f_i(z)/(z+1)$ are analytic at $z=1$ the
Mellin transforms (\ref{MEL1}) can be represented in terms of factorial series
\cite{FACSER}. Not all basic functions chosen above have this property. A 
corresponding analytic relation replacing
\begin{eqnarray}
f_i(z) \rightarrow f_i(1-z)
\end{eqnarray}
always exists. The additional terms are lower weight functions in $N$ or are
related to these by differentiation. We use this representation and consider 
the factorial series. Due to this both the pole--structure and the
asymptotic relation for $|N| \rightarrow \infty$ are known.~\footnote{In 
\cite{LILLE} 
asymptotic relations for non-alternating harmonic sums to low orders in 
$1/N^k$ were derived. Our algorithm given below is free of these 
restrictions. The main ideas were presented in January 2004 \cite{KITP}, see 
also \cite{LL2004}.}. The poles are located at the 
integers below a fixed value $N_0$. The recursion relations (\ref{MEL1}) are 
used to express the respective harmonic sums at any value $N~\in~{\bf C}$ 
except the poles.

Let us illustrate this representation in an example for the harmonic sum
$S_{2,1,1,1,1}(N)$. The corresponding basic function is
\begin{eqnarray}
\left(\frac{S_{1,4}(z)}{z-1}\right)_+~.
\end{eqnarray}
The recursion relation is given by
\begin{eqnarray}
\Mvec\left[\frac{S_{1,4}(z)}{z-1}\right](N+1)
&=& \hspace{3mm} \Mvec\left[\left(\frac{S_{1,4}(z)}{z-1}\right)_+\right](N) + 
\Mvec\left[S_{1,4}(z)\right](N)~, 
\end{eqnarray}
with
\begin{eqnarray}
\Mvec\left[S_{1,4}(z)\right](N) = \frac{1}{N+1}\left[\zeta_5 - \frac{1}{N+1} 
S_{1,1,1,1}(N)\right]~,
\end{eqnarray}
cf.~\cite{JBHK}.

The numerator function possesses a branch--point at $z=1$. The contributions
related to terms $\ln^k(1-z)/(z \pm 1)$ contained have to be subtracted 
explicitely due to there logarithmic growth (to a power) for $|N| \rightarrow 
\infty$. This is either possible using the relation  
$S_{1,4}(z)$ to $\Li_{5}(1-z)$
\begin{eqnarray}
S_{1,4}(z) &=& - \Li_5(1-z)+ \ln(1-z) \Li_4(1-z) - \frac{1}{2} \ln^2(1-z) 
\Li_3(1-z)  
+ \frac{1}{6} \ln^3(1-z) \Li_2(1-z) \nonumber\\  & &+ \frac{1}{24} \ln^4(1-z)  
\ln(z) + \zeta_5 
\end{eqnarray}
or considering harmonic sums, which are algebraic equivalent to the above and 
are related to a basic function which is regular at $z \rightarrow 1$. We 
will follow the latter way and use the algebraic relations \cite{ALGEBRA} to 
express $S_{2,1,1,1,1}(N)$ afterwards,
\begin{eqnarray}
S_{2,1,1,1,1} &=& S_{1,1,1,1,2} + \frac{1}{4} \Biggl[
                  S_1 S_{2,1,1,1} + S_{3,1,1,1,} + S_{2,2,1,1} +S_{2,1,2,1} +S_{2,1,1,2} 
\Biggr] 
\nonumber\\ 
&&
   -\frac{1}{12} \Biggl[S_1 
  S_{1,2,1,1} + S_{2,2,1,1} + S_{1,3,1,1}  
+ S_{1,2,1,2} - S_1 S_{1,1,2,1} - S_{2,1,2,1} 
- S_{1,1,3,1} - S_{1,1,2,2} \Biggr]
\nonumber\\ 
&&
  -\frac{1}{4} \Biggl[S_1 S_{1,1,1,2} + S_{2,1,1,2} +S_{1,2,1,2}
  + S_{1,1,2,2} + S_{1,1,1,3}\Biggr]
\end{eqnarray}
through known harmonic sums of lower weight. The latter sum obeys the 
representation 
\begin{eqnarray}
\label{eq11112}
S_{1,1,1,1,2}(N) &=& -\Mvec\left[\frac{\Li_5(1-x)}{1-x}\right](N) + \zeta_2 
S_{1,1,1,1}(N) - \zeta_3 S_{1,1,1}(N) \nonumber\\
& & + \zeta_4 S_{1,1}(N) - \zeta_5 S_1(N) +\zeta_6~.
\end{eqnarray}
The function in the remaining Mellin transform is regular at $z=1$ and can be 
represented in terms of a factorial series. The remainder terms in 
(\ref{eq11112}) are polynomials of single harmonic sums. Therefore the poles 
of $S_{1,1,1,1,2}(N)$, resp. $S_{2,1,1,1,1}(N)$, are located at the 
non-positive integers. Finally we need the asymptotic representations of 
$\Mvec\left[{\Li_5(1-x)}/{(1-x)}\right](N)$, 
\begin{eqnarray}
\Mvec\left[\frac{\Li_5(1-x)}{1-x}\right](z) &\sim&
\frac{1}{z}+\frac{1}{32 z^2}-{\frac 
{179}{7776}}\,\frac{1}{z^3}+{\frac {515}{
41472}}\,\frac{1}{z^4}-{\frac 
{216383}{194400000}}\,\frac{1}{z^5}-{\frac {183781
}{25920000}}\,\frac{1}{z^6}
\nonumber\\ & &
+{\frac 
{4644828197}{653456160000}}\,\frac{1}{z^7}
+{
\frac {153375307}{49787136000}}\,\frac{1}{z^8}-{\frac 
{371224706507}{
25204737600000}}\,\frac{1}{z^9}
\nonumber\\ & &
+{\frac 
{959290541}{160030080000}}\,\frac{1}{z^{10}}
+{\frac 
{575134377343021}{16913534146740000}}\,\frac{1}{z^{11}}
-{\frac {
14855426650259}{312400053504000}}\,\frac{1}{z^{12}}
\nonumber\\ & &
-{\frac {
29106619674489691525729}{319702820637227227200000}}\,\frac{1}{z^{13}}+{\frac
{225456132288901603}{788601079506240000}}\,\frac{1}{z^{14}}
\nonumber\\ & &
+{\frac 
{
263567702701300558681}{1053965342760089760000}}\,\frac{1}{z^{15}}-{\frac 
{
355061945309358701}{187184432058624000}}\,\frac{1}{z^{16}}
\nonumber\\ &&
-{\frac 
{
1432477558547377054456843733}{4988266898917709221214400000}}\,\frac{1}{z^{17}}
\nonumber\\ &&
+{\frac 
{192140702840923335916939}{13028192458306945920000}}\,\frac{1}{z^{18}}
\nonumber\\ &&
-{\frac {2027981189268747465011536794768001}{
254294408120596135866406712880000}}\,\frac{1}{z^{19}}
+ O\left(\frac{1}{z^{20}}\right)
\end{eqnarray}
The corresponding representations for all other harmonic sums of weight 
{\sf w = 6} will be given in a forthcoming paper.
\section{Conclusions}
%

\vspace{1mm}
\noindent
We derived the basic functions spanning the nested harmonic (alternating) 
sums up to weight {\sf w = 6} with no index $\{-1\}$. This sub-class governs
the functions contributing to the massless single-scale quantities, like
the anomalous dimensions and Wilson coefficients to 3-loop order in QED and 
QCD. There are first indications, that in the massive case, even in the limit 
$Q^2 \gg m^2$ this class needs to be extended at 3-loop order, cf. \cite{HEAV6}. 
Up to weight {\sf w = 5} all basic functions were given by polynomials of 
Nielsen integrals, Eq.~(\ref{NIEL1}), of argument $x$ or $-x$ weighted 
by $1/(x \pm 1)$. 
Although most of the basic functions
at {\sf w = 6} share this property, some contain 1-dimensional integrals over
polynomials of Nielsen integrals $\left. A_{i}(\pm x)\right|_{1 ... 3}$ and 
more dimensional integrals, which are not reducible. This is generally expected and
the cases up to {\sf w = 5} form an exception.

We lined out how the exact the representation of the Mellin transforms of the 
basic functions can be obtained, generalizing effective numerical high-precision  
representations \cite{ANCONT1,AB}. Up to terms which can be determined 
algebraically the Mellin transforms of the basic functions are factorial 
series. The singularities of the Mellin transforms are located at the 
non-positive integers. They obey recursion relations for $N \rightarrow N+1$.
The asymptotic representation of the Mellin transforms can be determined
analytically. The basic Mellin transforms are thus generalizations of Euler's
$\psi$-function and their derivatives, which describe the single harmonic 
sums.

\vspace{3mm}\noindent
{\bf Acknowledgment.}\\
I would like to thank Steve Rosenberg for his invitation and the organization 
of a very interesting and stimulating workshop. For useful conversations I would 
like to thank J. Ablinger, Y. Andre, F.~Brown, and C. Schneider. This paper 
was supported in part by the Clay Mathematics Institute, DFG 
Sonderforschungsbereich Transregio~9, Computergest\"utzte Theoretische Physik, 
and by the European Commission MRTN HEPTOOLS under Contract No. MRTN-CT-2006-035505.

\newpage
\section{Appendix~A: Useful Integrals}
%

\vspace{1mm}
\noindent
In this appendix we list useful constants and integrals.
\begin{eqnarray}
\Li_{k}(1) &=& \zeta_k 
\\
S_{1,k}(1) &=& \zeta_{k+1} 
\\
S_{2,2}(1) &=& \frac{1}{10} \zeta_2^2  
\\
S_{3,2}(1) &=& 2 \zeta_5 - \zeta_2 \zeta_3 
\\
S_{3,2}(-1) &=& -\frac{29}{32} \zeta_5 + \frac{1}{2} \zeta_2 \zeta_3 
\\
S_{2,3}(1) &=& 2 \zeta_5 - \zeta_2 \zeta_3 
\\
A_1(1) &=& -3 \zeta_5 + 2 \zeta_2 \zeta_3 
\\
A_1(-1) &=& -\frac{17}{16} \zeta_5 + \frac{3}{4} \zeta_2\zeta_3 
\\
A_2(1) &=& -\frac{1}{2} \zeta_5  
\\
A_3(1) &=& -3 \zeta_5 +\zeta_2 \zeta_3 
\end{eqnarray}

\begin{eqnarray}
\int_0^x dy \frac{\Li_3(-y)}{1+y} &=& \ln(1+x) \Li_3(-x) + \frac{1}{2}
\Li_2^2(-x) 
\\
\int_0^x dy \frac{\ln(y) \Li_2(-y)}{1+y} &=& \ln(1+x) \ln(x) \Li_2(-x) 
+ \frac{1}{2} \Li_2^2(-x) 
\nonumber\\ 
&&
- 2 S_{2,2}(-x) + 2 \ln(x) S_{1,2}(-x)
\\
\int_0^x dy \frac{S_{1,2}(y)}{y-1} &=& \ln(1-x) S_{1,2}(x) + 3 S_{1,3}(x)
\\
\int_0^x \frac{dy}{y} \left[\Li_2(1-y)- \zeta_2\right]
&=& - 2 \Li_3(x) + \ln(x) \Li_2(x)
\\
\int_0^x dy \frac{\Li_3(y)}{y-1} &=& \frac{1}{2} \Li_2^2(x) + \ln(1-x) 
\Li_3(x)
\\
\int_0^x dy \frac{\ln(y)}{y-1} \Li_2(y) &=&
\frac{1}{2} \Li^2_2(x) + \ln(x)\ln(1-x) \Li_2(x) \nonumber\\ &&
- 2 S_{2,2}(x)
+ 2 \ln(x) S_{1,2}(x)
\\
\int_0^x \frac{dy}{y} \ln(1-y) \Li_3(y) &=& - \Li_2(x) \Li_3(x) + A_1(x)
\\
\int_0^x \frac{dy}{y} 
\Li_2(y) \ln(y) \ln(1-y) &=& - \frac{1}{2} \ln(x) \Li_2^2(x) 
+ \frac{1}{2} A_1(x) 
\\
\int_0^x \frac{dy}{y} 
\Li_2(-y) \ln(y) \ln(1+y) &=& - \frac{1}{2} \ln(x) \Li_2^2(-x) 
+ \frac{1}{2} A_1(-x) 
\\
\int_0^x \frac{dy}{y} 
\Li_3(-y) \ln(1+y)  &=& - \Li_2(-x) \Li_3(-x)  
+ A_1(-x) 
\\
\int_0^x \frac{dy}{y} \left[\Li_3(1-x) - \zeta_3\right] &=&
- S_{1,2}(x) \ln(x)  + \frac{1}{2} \Li_2^2(x) -\zeta_2 \Li_2(x) 
\end{eqnarray} \begin{eqnarray}
\int_0^x \frac{dy}{y} S_{1,2}(y) \ln(y) &=& S_{2,2}(x) \ln(x) - S_{3,2}(x)
\\
\int_0^x \frac{dy}{y} S_{1,2}(-y) \ln(y) &=& S_{2,2}(-x) \ln(x) - S_{3,2}(-x)
\end{eqnarray}

\newpage

\end{document}